%% file: main.tex
\documentclass[12pt,onecolumn]{iopjournal}

\usepackage[T1]{fontenc}
\usepackage[utf8]{inputenc}
\usepackage[english]{babel}
\usepackage{amsmath,amssymb}
\usepackage{esint}
\usepackage{physics}
\usepackage{placeins}
\usepackage{microtype}
\usepackage{newtxtext,newtxmath}
\usepackage{ragged2e}
\usepackage{cite}
\fancyhfoffset[L]{0pt}
\renewcommand{\footrulewidth}{0pt}
\hypersetup{pdftitle={Nonlinear optical response with relaxation in the velocity gauge},
  pdfauthor={Carlos M. Pedrosa et al.},hidelinks}

\renewcommand{\articletype}[1]{%
  \noindent{\Large\sffamily Journal of Physics: Condensed Matter}\par
  \vspace{6mm}%
  \noindent{\scriptsize\sffamily\bfseries\MakeUppercase{#1}}\par
  \vspace{5mm}}
\AtBeginDocument{\justifying}

\begin{document}
\articletype{Paper}
\raggedbottom
\title{Nonlinear optical response with relaxation in the velocity gauge}
\author{ Carlos M. Pedrosa$^{1,2}$, João P. C. Ferreira$^{1,2}$, João Pinho$^{1,2}$, Yu. V. Bludov$^{1,3}$, B. Amorim$^{1,2}$ and João M. Viana Parente Lopes$^{1,2}$}
\par\vspace{2mm}
\affil{$^1$Physics Center of Minho and Porto Universities (CF-UM-UP), Laboratory of Physics for Materials and Emergent Technologies (LaPMET)}\par
\affil{$^2$Faculty of Science, University of Porto, Campo Alegre, 4169-007 Porto, Portugal}\par
\affil{$^3$Department of Physics, University of Minho, Campus of Gualtar, 4710-057 Braga, Portugal}\par\medskip

\selectlanguage{english}%
\begin{abstract}
We develop a framework for the evaluation of the nonlinear optical conductivity of electrons in crystals, driven by homogeneous optical fields, taking into account dissipation effects within the relaxation time approximation (RTA). The present formalism is based on Hamiltonians with minimal light-matter coupling within the velocity gauge, where the electric field is expressed in terms of  a vector potential. The nonlinear conductivities are obtained by perturbatively solving the equation of motion for the single-particle reduced density matrix, which includes dissipative effects via the relaxation time approximation. We present a perturbative method to evaluate the reference equilibrium density matrix, towards which dissipation relaxes the system, which in the velocity gauge adiabatically follows the applied vector potential, and is also broadened by dissipative effects. Comparing with previous formulations based on the length gauge, the present approach has the benefit of avoiding the cumbersome evaluation of derivatives of Bloch states with respect to momentum. We test the presented method against length-gauge results, and calculate second- and third-order responses in a graphene-like tight-binding model with a staggered sublattice potential and a periodic modulation. We find that the relaxation time approximation gives results that deviate both quantitatively and qualitatively from the simple inclusion of an imaginary damping term in the frequencies, referred to as adiabatic switching.  At equal relaxation parameter, the leading above-gap injection contribution differs by a factor of two; below the gap, the numerical results indicate a finite imaginary dc response in the small-relaxation limit of the relaxation-time calculation. These results show that adiabatic switching does not generally reproduce relaxation through the equation of motion, even in the small-relaxation limit.
\end{abstract}

\input{introduction}

\section{General Formalism}
\label{sec:formalism}

\subsection{Free Hamiltonian and conventions}

We consider a non-interacting electronic system with translational
symmetry, which may describe either a lattice model or electrons
moving in a continuum periodic potential. In either case, the
single-particle Hamiltonian admits a Bloch representation.
In a basis of localized orbitals, the Hamiltonian can be
written as
\begin{equation}
    H_0 = \sum_{\vb R,\vb R'}\sum_{\alpha,\beta}
    t_{\alpha\beta}(\vb R-\vb R')
    c^\dagger_{\vb R\alpha}c_{\vb R'\beta},
    \label{eq:real_space_H}
\end{equation}
where $\alpha,\beta$ label the orbital types, $\vb R$ and
$\vb R'$ denote the positions of the unit cells, and
$c^\dagger_{\vb R\alpha}$ and $c_{\vb R\alpha}$
are the corresponding creation and annihilation operators.
The coefficients $t_{\alpha\beta}(\vb R-\vb R')$
are matrix elements of the single-particle Hamiltonian
and are not restricted to nearest-neighbor hopping.
For continuum systems, such a representation can be obtained
by expanding the Hamiltonian in a Wannier basis, which may
be taken to be complete if no band truncation is performed.

Denoting by $\vb r_\alpha$ the position of orbital $\alpha$
relative to the origin of its unit cell, we introduce
the reciprocal-space operators through the Fourier convention
\begin{align}
    c_{\vb k\alpha}
    &= \frac{1}{\sqrt N}\sum_{\vb R}
    c_{\vb R\alpha}
    e^{-i\vb k\cdot(\vb R+\vb r_\alpha)},\\
    c_{\vb R\alpha}
    &= \frac{1}{\sqrt N}\sum_{\vb k\in\mathrm{FBZ}}
    c_{\vb k\alpha}
    e^{i\vb k\cdot(\vb R+\vb r_\alpha)},
    \label{eq:Fourier_conv}
\end{align}
where $N$ is the number of unit cells.
With this convention, the Hamiltonian becomes
\begin{equation}
    H_0 = \sum_{\vb k\in\mathrm{FBZ}}
    \sum_{\alpha,\beta}
    (H_{0,\vb k})_{\alpha\beta}
    c^\dagger_{\vb k\alpha}c_{\vb k\beta},
\end{equation}
where
\begin{equation}
    (H_{0,\vb k})_{\alpha\beta}
    = \sum_{\vb R}
    t_{\alpha\beta}(\vb R)
    e^{-i\vb k\cdot(\vb R+\vb r_\alpha-\vb r_\beta)}.
    \label{eq:bloch_H}
\end{equation}

The inclusion of the intracell orbital positions in the
Fourier phases is particularly convenient when describing
the coupling to a spatially uniform electromagnetic field.
For models in which the position operator is diagonal in
the localized-orbital basis, this convention allows the
light--matter coupling to be implemented through a simple
shift of crystal momentum. However, in a general Wannier
representation, the position operator may also have
off-diagonal matrix elements, which must be accounted for
when constructing the electromagnetic coupling.
This distinction does not depend on whether the underlying
system is described by a lattice or continuum Hamiltonian.

\subsection{Perturbative series of the Hamiltonian and velocity}

In the velocity gauge, a spatially uniform electric field
$\vb E(t)=-\partial_t\vb A(t)$ is represented by a
time-dependent vector potential $\vb A(t)$.
We denote by $H_{\vb k}(\vb A)$ the corresponding
single-particle Hamiltonian, expressed in a fixed Bloch
basis, and define
\begin{equation}
    H_{\vb k}(t)
    \equiv H_{\vb k}(\vb A(t))
    =H_{0,\vb k}+V_{\vb k}(t),
    \label{eq:field_hamiltonian}
\end{equation}
where $H_{0,\vb k}=H_{\vb k}(\vb A=0)$.
Assuming an expansion in powers of the vector potential,
the perturbation can be written as
\begin{align}
V_{\vb k}(t)
&=\sum_{n=1}^{\infty}\frac{(-q)^n}{n!}
A_{j_1}(t)\cdots A_{j_n}(t)\,
h^{j_1\cdots j_n}_{\vb k},
\label{eq:perturbation_expansion}
\end{align}
where $q=-e$ is the electron charge, repeated Cartesian
indices are implicitly summed, and
\begin{equation}
h^{j_1\cdots j_n}_{\vb k}
=
\frac{1}{(-q)^n}
\left.
\frac{\partial^n H_{\vb k}(\vb A)}
{\partial A_{j_1}\cdots\partial A_{j_n}}
\right|_{\vb A=0}
\label{eq:general_coupling_tensor}
\end{equation}
defines the $n$th-order light-matter coupling tensor.

The field-dependent velocity operator is obtained from
the derivative of the Hamiltonian with respect to the
vector potential,
\begin{equation}
v^i_{\vb k}(t)
=
-\frac{1}{q}
\left.
\frac{\partial H_{\vb k}(\vb A)}
{\partial A_i}
\right|_{\vb A=\vb A(t)}.
\label{eq:general_velocity}
\end{equation}
It therefore admits the expansion
\begin{align}
v^i_{\vb k}(t)
&=\sum_{n=0}^{\infty}\frac{(-q)^n}{n!}
A_{j_1}(t)\cdots A_{j_n}(t)\,
h^{i j_1\cdots j_n}_{\vb k}.
\label{eq:field_velocity}
\end{align}
In particular, $h^i_{\vb k}$ is the $i$th Cartesian
component of the unperturbed velocity operator.

For the class of models considered in the applications
below, the coupling to a uniform vector potential is
implemented through the momentum shift
\begin{equation}
    H_{\vb k}(\vb A)
    =H_{0,\vb k-\frac{q}{\hbar}\vb A}.
    \label{eq:minimal_coupling}
\end{equation}
This prescription is exact for continuum Hamiltonians
with conventional minimal coupling in a fixed
cell-periodic Bloch representation. It also follows
from the Peierls substitution for lattice models
with a diagonal position operator in the localized-orbital
basis, where the hopping amplitudes transform as $t_{\alpha\beta}(\vb R)
\rightarrow
t_{\alpha\beta}(\vb R)
e^{i\frac{q}{\hbar}\vb A(t)\cdot
(\vb R+\vb r_\alpha-\vb r_\beta)}
\label{eq:peierls_hopping}$. When Eq.~\eqref{eq:minimal_coupling} holds, the coupling
tensors reduce to ordinary momentum derivatives,
\begin{align}
h^{j_1\cdots j_n}_{\vb k}
&=\frac{1}{\hbar^n}
\frac{\partial^n H_{0,\vb k}}
{\partial k_{j_1}\cdots\partial k_{j_n}},
\label{eq:coupling_tensor}
\end{align}
and the field-dependent velocity can equivalently
be expressed as
\begin{equation}
v^i_{\vb k}(t)
=
\frac{1}{\hbar}
\frac{\partial H_{\vb k}(t)}{\partial k_i}.
\label{eq:momentum_velocity}
\end{equation}

More generally, Eq.~\eqref{eq:minimal_coupling} need
not hold in an arbitrary Wannier or band representation.
For truncated band or Wannier Hamiltonians, additional care
is required to construct the electromagnetic coupling
consistently with the projection. The corresponding
velocity matrix elements are obtained through
covariant momentum derivatives, which include
the Berry connection associated with the chosen
basis~\cite{ventura_gauge_2017,passos_nonlinear_2018}.

Importantly, the perturbative formalism developed
in this work does not rely on Eq.~\eqref{eq:coupling_tensor}.
Instead, the tensors defined in
Eq.~\eqref{eq:general_coupling_tensor} enter the
density-matrix recursion as the coefficients of the
field-dependent Hamiltonian. Thus, while their
explicit construction may involve Berry connections
or covariant derivatives of Bloch states for truncated Wannier models, the resulting perturbative
equations do not require momentum derivatives
acting on the density matrix, even in the presence
of relaxation.

\subsection{Current density expansion}
Both the velocity operator and the reduced density matrix must 
be expanded to obtain the nonlinear current. The expectation value of the
current density is
\begin{align}
\expval{J_i}(t)=q\int\frac{d^d\vb k}{(2\pi)^d}\sum_{ss'}
 v^{i}_{\vb kss'}(t)\,\rho_{\vb ks's}(t),
\label{eq:av_current}
\end{align}
where $d$ is the spatial dimension and $s,s'$ label Bloch bands. The
single-particle reduced density matrix is defined by
\begin{align}
\rho_{\vb k,s's}(t)
&=\Tr\!\left[c^{\dagger}_{\vb ks}c_{\vb ks'}\hat\varrho(t)\right],
\label{eq:reduced_density_matrix}
\end{align}
with $c^{\dagger}_{\vb ks}$ and $c^{}_{\vb ks}$ the creation and annihilation
operators for a Bloch state labelled $s$, and $\hat\varrho(t)$ the many-body density
operator.

With the Fourier convention $X(t)=\int d\omega\,e^{-i\omega t}X(\omega)/(2\pi)$,
the current is expanded in powers of the electric field as
\begin{align}
\expval{J_i}(\omega)
&=\sum_{n=1}^{\infty}\expval{J_i}^{(n)}(\omega).
\end{align}
For compactness, we denote the ordered tuple of input frequencies by
$\omega_{1:n}\equiv(\omega_1,\ldots,\omega_n)$,
with $d^n\omega_{1:n}\equiv\prod_{\ell=1}^n d\omega_\ell$
and $\omega_{1\cdots n}\equiv\sum_{\ell=1}^n\omega_\ell$.
The $n$th-order conductivity tensor
$\sigma_{i j_1\cdots j_n}(\omega_{1:n})$ is defined by
\begin{align}
\expval{J_{i}}^{(n)}(\omega)
&=\int\frac{d^n\omega_{1:n}}{(2\pi)^{n-1}}\,
\sigma_{i j_1\cdots j_n}(\omega_{1:n})
\delta(\omega-\omega_{1\cdots n})
E_{j_1}(\omega_1)\cdots E_{j_n}(\omega_n).\label{cond_definition}
\end{align}
\subsection{Equation of motion}
We account for dissipation using a relaxation-time approximation in
the equation of motion for the reduced density matrix,
\begin{align}
i\hbar\pdv{\rho_{\vb k}}{t}
&=[H_{\vb k}(t),\rho_{\vb k}]
-i\gamma\bigl(\rho_{\vb k}-\rho^{\rm eq}_{\vb k}(t)\bigr),
\label{eq:RTA}
\end{align}
where $\gamma$ is the relaxation energy ($\tau=\hbar/\gamma$ is the
corresponding relaxation time) and $\rho^{\rm eq}_{\vb k}(t)$ is the
field-dependent equilibrium density matrix. This reference state is a function of the instantaneous Hamiltonian
and therefore carries the same vector-potential dependence. An equation of this form can be obtained from an
open-system description with B\"uttiker probes in the slow-driving,
wide-band limit~\cite{pinho_exact_2026}. Coupling to these Büttiker probes also leads to a broadening of the single-particle energy levels in the equilibrium state. This can be taken into account via the Lorentzian-broadened Fermi--Dirac function
\begin{align}
f(\epsilon)
&=\int_{-\infty}^{\infty}\!\frac{d\epsilon'}{\pi}\,
 \frac{\gamma\,f_{\rm FD}(\epsilon')}
 {(\epsilon-\epsilon')^2+\gamma^2},
\label{eq:broadened_fermi}
\end{align}
where $f_{\rm FD}(\epsilon)$ is the Fermi--Dirac distribution,
\begin{align}
    f_{\rm FD}(\epsilon)=\frac{1}{1+e^{\beta(\epsilon-\mu)}},
\end{align}
with $\beta$ being the inverse temperature and $\mu$ the Fermi energy. Here we use the same parameter $\gamma$ for the Lorentzian half-width
and the relaxation energy in Eq.~\eqref{eq:RTA}.
\subsection{Perturbation Theory}
To solve the equation of motion, we introduce the deviation from the
instantaneous equilibrium state,
\begin{align}
\Delta\rho_{\vb k}(t)
&=\rho_{\vb k}(t)-\rho^{\rm eq}_{\vb k}(t).
\end{align}
Because $\rho^{\rm eq}_{\vb k}(t)$ is a function of $H_{\vb k}(t)$,
it commutes with the instantaneous Hamiltonian. Consequently,
\begin{align}
i\hbar\pdv{\Delta\rho_{\vb k}}{t}
&=[H_{0,\vb k},\Delta\rho_{\vb k}]
-i\gamma\Delta\rho_{\vb k}
+[V_{\vb k},\Delta\rho_{\vb k}]
-i\hbar\pdv{\rho^{\rm eq}_{\vb k}}{t}.
\label{eq:delta_rho_motion}
\end{align}

We solve Eq.~\eqref{eq:delta_rho_motion} recursively in frequency
space. Writing $\Delta\rho_{\vb k}=\sum_{n=1}^{\infty}\Delta\rho^{(n)}_{\vb k}$,
with $\Delta\rho^{(n)}_{\vb k}$ of order $n$ in the vector potential, we consider a perturbative treatment, so that $\Delta\rho_{\vb{k}}^{(n+1)}\ll\Delta\rho_{\vb{k}}^{(n)}$.
We define the expansion coefficients by 
\begin{align}
\Delta\rho^{(n)}_{\vb kss'}(\omega)
&=(-q)^n\int\!\frac{d^n\omega_{1:n}}{(2\pi)^{n-1}}
\Delta\rho^{j_1\cdots j_n}_{\vb kss'}(\omega_{1:n})\,
A_{j_1}(\omega_1)\cdots A_{j_n}\,
(\omega_n)\,
\delta(\omega-\omega_{1\cdots n}),
\label{isolated_orders_rho}
\end{align}
where the coefficients of the expansion obey the recursive relation 
\begin{align}\label{coefs_delta_rho}
\Delta \rho^{j_{1}\cdots j_{n}}_{\vb kss'}\left(\omega_{1},\dots,\omega_{n}\right)  &=\frac{1}{\hbar\omega_{1\cdots n}+i\gamma-\Delta\epsilon_{\vb kss'}}\times  \\
&\quad \left\{ \sum^{n}_{m=1}\frac{1}{m!}\left[h^{j_{1}\cdots j_{m}}_{\vb k},\Delta\rho^{j_{m+1}\cdots j_{n}}_{\vb k}\left(\omega_{m+1:n}\right)\right]_{ss'}-\hbar\omega_{1\cdots n}\rho^{j_{1}\cdots j_{n}}_{{\rm eq},\vb kss'}\right\},\nonumber
\end{align}
with  $\Delta\epsilon_{\vb kss'}\equiv
\epsilon_{\vb ks}-\epsilon_{\vb ks'}$ being band-energy difference in the eigenbasis of the unperturbed Hamiltonian $H_{0,\vb k}\ket{s}=\epsilon_{\vb ks}\ket{s}$. Note that in the equation above we also introduced the notation $\omega_{m+1:n}=(\omega_{m+1},\dots,\omega_{n})$. The coefficient $\rho^{j_1\cdots j_n}_{{\rm eq},\vb k}$
is defined by expanding $\rho^{\rm eq}_{\vb k}(t)$ in the
same way as $\Delta\rho_{\vb k}(t)$ in
Eq.~\eqref{isolated_orders_rho}. Its time derivative produces
the final term in Eq.~\eqref{coefs_delta_rho}.
\subsection{Conductivity tensor}
The instantaneous-equilibrium contribution to the current
vanishes upon integration over the full Brillouin zone.
Indeed, with $F'(x)=f(x)$, the identity
\begin{equation}
\tr[v^i_{\vb k}f(H_{\vb k})]
=\frac{1}{\hbar}\partial_{k_i}
\operatorname{tr}F(H_{\vb k})
\end{equation}
makes this contribution a total momentum derivative,
whose integral vanishes by Brillouin-zone periodicity.
Thus only the deviation $\Delta\rho_{\vb k}$ contributes
to the net current. With this in mind, inserting Equation (\ref{isolated_orders_rho}) into (\ref{eq:av_current})
yields the nonlinear conductivity tensor
\begin{align}
\tilde\sigma_{i j_{1}\cdots j_{n}}(\omega_{1:n})=\frac{i^{n}q^{n+1}}{\omega_{1}\cdots\omega_{n}}\int\frac{d^{d}\vb k}{(2\pi)^{d}}\sum^{n-1}_{m=0}\frac{1}{m!} \tr\left[h^{i j_{1}\cdots j_{m}}_{\vb k}\Delta\rho^{j_{m+1}\cdots j_{n}}_{\vb k}(\omega_{m+1:n})\right].
\label{conds_general}
\end{align}
The trace is over band indices. For reference we write here the first order expression,
\begin{align}
\tilde{\sigma}_{i j_1}(\omega_1)=-i\hbar q^2\int_{\mathrm{FBZ}}\frac{\mathrm d^d\vb{k}}{(2\pi)^d}\sum_{ss'}\frac{h^i_{\mathbf k,ss'}\rho^{j_1}_{\mathrm{eq},\mathbf k,s's}}{\hbar\omega_1+i\gamma-\Delta\epsilon_{\mathbf k,s's}},
\end{align}
and the second order expression,
\begin{align}
\begin{split}
    &\tilde{\sigma}_{i j_1j_2}(\omega_{1},\omega_2)
=\frac{\hbar q^3}{\omega_1\omega_2}\int_{\mathrm{FBZ}}\frac{\mathrm d^d \vb{k}}{(2\pi)^d}
\sum_{ss'}\Bigg\{\frac{h^i_{\mathbf k,ss'}}{\hbar\omega_{12}+i\gamma-\Delta\epsilon_{\mathbf k,s's}}\Bigg[\omega_{12}\rho^{j_1j_2}_{\mathrm{eq},\mathbf k,s's}\\
&\qquad+\omega_2\sum_r\Bigg(\frac{h^{j_1}_{\mathbf k,s'r}\rho^{j_2}_{\mathrm{eq},\mathbf k,rs}}{\hbar\omega_2+i\gamma-\Delta\epsilon_{\mathbf k,rs}}
-\frac{\rho^{j_2}_{\mathrm{eq},\mathbf k,s'r}h^{j_1}_{\mathbf k,rs}}{\hbar\omega_2+i\gamma-\Delta\epsilon_{\mathbf k,s'r}}\Bigg)\Bigg]
+\frac{\omega_2 h^{i j_1}_{\mathbf k,ss'}\rho^{j_2}_{\mathrm{eq},\mathbf k,s's}}{\hbar\omega_2+i\gamma-\Delta\epsilon_{\mathbf k,s's}}\Bigg\},
\end{split}
\end{align}
where all the sums are over band indices. To obtain the corresponding adiabatic-switching expressions, we omit the relaxation term from the equation of motion, and take the initial equilibrium occupations to be given by the unbroadened Fermi--Dirac distribution. In practice, one only has to first take $\gamma=0$ in the expressions above, use $f=f_{\mathrm{FD}}$ in the equilibrium-density coefficients, and subsequently apply the frequency shifts $\hbar\omega\to\hbar\omega+i\gamma$ to each input frequency. In particular, at second order, the denominator associated with the frequency $\omega_{12}$ becomes $\hbar\omega_{12}+2i\gamma-\Delta\epsilon_{\mathbf kss'}$, instead of $\hbar\omega_{12}+i\gamma-\Delta\epsilon_{\mathbf kss'}$ in the relaxation-time approximation. Consequently, the two prescriptions generally yield different nonlinear conductivities at finite $\gamma$.

Because the electric-field factors in Eq.~(\ref{cond_definition}) commute, the physical conductivity tensor must be symmetric under simultaneous permutations of the field indices and their associated frequencies. Since this symmetry is not manifest in Eq.~(\ref{conds_general}), we introduce the permutation-symmetric conductivity tensor,
\begin{align}
    \sigma_{i j_1\cdots j_n}(\omega_1,\dots,\omega_n) = \frac{1}{n!}\sum_{\pi\in S_n}\tilde{\sigma}_{i j_{\pi(1)}\cdots j_{\pi(n)}}(\omega_{\pi(1)},\dots,\omega_{\pi(n)}),
\end{align}
where we sum over all permutations $\pi\in S_n$ of the $n$ field index-frequency pairs. This is the quantity that appears in the figures throughout this work.

The remaining task is therefore the computation of the equilibrium-density
coefficients $\rho^{j_{1}\cdots j_{n}}_{{\rm eq},\vb k}$ entering
Eq.~(\ref{coefs_delta_rho}), which are independent of the input frequencies. Their systematic construction is
developed in the next section.

\section{Expansion of the Equilibrium Density Matrix}
\label{sec:equilibrium}
\subsection{Expansion in $V_{\vb k}(t)$}
Equation~(\ref{coefs_delta_rho}) depends on the expansion coefficients
of the adiabatically evolved equilibrium density matrix. We now derive
a systematic procedure for obtaining these coefficients to arbitrary
order. The equilibrium density matrix is
\begin{equation}
\rho^{\rm eq}_{\vb k}(t)=f\!\left(H_{\vb k}(t)\right).
\end{equation}
Here, $f$ is the Lorentzian-broadened occupation of
Eq.~\eqref{eq:broadened_fermi}, and
$H_{\vb k}(t)=H_{0,\vb k}+V_{\vb k}(t)$.

Following Appendix \ref{app:deriv_expansion}, we  express this matrix as $\rho^\text{eq}_{\vb k}(t)=\sum_{l=0}^\infty\rho^{[l]}_{\vb k}(t)$, and the terms $\rho^{[l]}_{\vb k}(t)$, containing $l$ insertions of the potential $V_{\vb k}(t)$, have the following expansion: 
\begin{align}
\rho^{[l]}_{\vb k,s_0s_l}(t)
&=\sum_{s_1,\ldots,s_{l-1}}
 V_{\vb k,s_0s_1}(t)\cdots
 V_{\vb k,s_{l-1}s_l}(t)\,
\mathcal C_{s_0\cdots s_l},\label{eq:expansion_potentialV}
\end{align}
where the dependence on the equilibrium distribution is contained in
the permutation-symmetric coefficient below:
\begin{equation}
\mathcal C_{s_0\cdots s_l}
=\sum_{j=0}^{l}
\frac{f(\epsilon_{s_j})}
 {\displaystyle\prod_{\ell\ne j}
 (\epsilon_{s_j}-\epsilon_{s_\ell})}.
\label{eq:large_S}
\end{equation}
This coefficient constitutes the basic building block of the equilibrium-density
expansion and is valid whenever the energies are distinct; the cases for coincident
energies $\epsilon_{s_j},\epsilon_{s_\ell}$ are handled by taking the corresponding continuous limit $\epsilon_{s_j}\to\epsilon_{s_\ell}$,
which involves derivatives of $f$.

Equation (\ref{eq:large_S}) is one of the central results of this work. It achieves two main objectives: it shows that the coefficient $\mathcal C_{s_0\cdots s_l}$ is written solely in terms of the unperturbed eigenenergies and
the occupation function --- eliminating the need to explicitly
evaluate the covariant derivative acting on the density matrix, as was initially
believed necessary \cite{passos_nonlinear_2018} --- and it provides a systematic
way to incorporate bath broadening into the initial equilibrium state through the occupation function $f$ defined in
Eq.~\eqref{eq:broadened_fermi}, without derivatives of Bloch eigenvectors. Finally, although derived for a Bloch Hamiltonian, this
matrix-function identity also holds for a general finite-dimensional
Hermitian Hamiltonian.
\subsection{First-order}
To illustrate the result, let $f_s\equiv f(\epsilon_{\vb ks})$.
The first-order equilibrium-density coefficient is
\begin{equation}
\rho^j_{{\rm eq},\vb kss'}
=h^j_{\vb k,ss'}
\frac{f_s-f_{s'}}{\epsilon_s-\epsilon_{s'}}.
\label{eq:eq_rho_first}
\end{equation}
For $\epsilon_s=\epsilon_{s'}$, the ratio becomes
$f'(\epsilon_s)$, including the cases of degenerate bands.

\subsection{Second-order}
At second order, one contribution arises directly from
$h^{j_1j_2}_{\vb k}$, while the other involves two
first-order couplings and a sum over intermediate bands:
\begin{align}
\rho^{j_1j_2}_{{\rm eq},\vb kss'}
&=\frac12 h^{j_1j_2}_{\vb k,ss'}
\frac{f_s-f_{s'}}{\epsilon_s-\epsilon_{s'}}
+\sum_{s_1}h^{j_1}_{\vb k,ss_1}
 h^{j_2}_{\vb k,s_1s'}\,
 \mathcal C_{ss_1s'}.
\label{eq:eq_rho_second}
\end{align}
Here $\mathcal C_{ss_1s'}$ is computed by plugging $s_0=s$ and $s_2=s'$ into:
\begin{equation}
    \mathcal C_{s_0s_1s_2}=\frac{\epsilon_{s_0}\left(f_{s_2}-f_{s_1}\right)+\epsilon_{s_1}\left(f_{s_0}-f_{s_2}\right)+\epsilon_{s_2}\left(f_{s_1}-f_{s_0}\right)}{(\epsilon_{s_0}-\epsilon_{s_1})(\epsilon_{s_0}-\epsilon_{s_2})(\epsilon_{s_1}-\epsilon_{s_2})},
\end{equation}
obtained by developing the summations and products in Eq.~\eqref{eq:large_S}; and then taking the proper energy
limits for coinciding energies, as explained above. 
At arbitrary order, these equilibrium-density coefficients
enter the recursion in Eq.~\eqref{coefs_delta_rho}, yielding the
conductivity in Eq.~\eqref{conds_general}.

\section{Numerical results}
\label{sec:results}

\begin{figure}[h]
\centering
\includegraphics[width=0.9\linewidth]{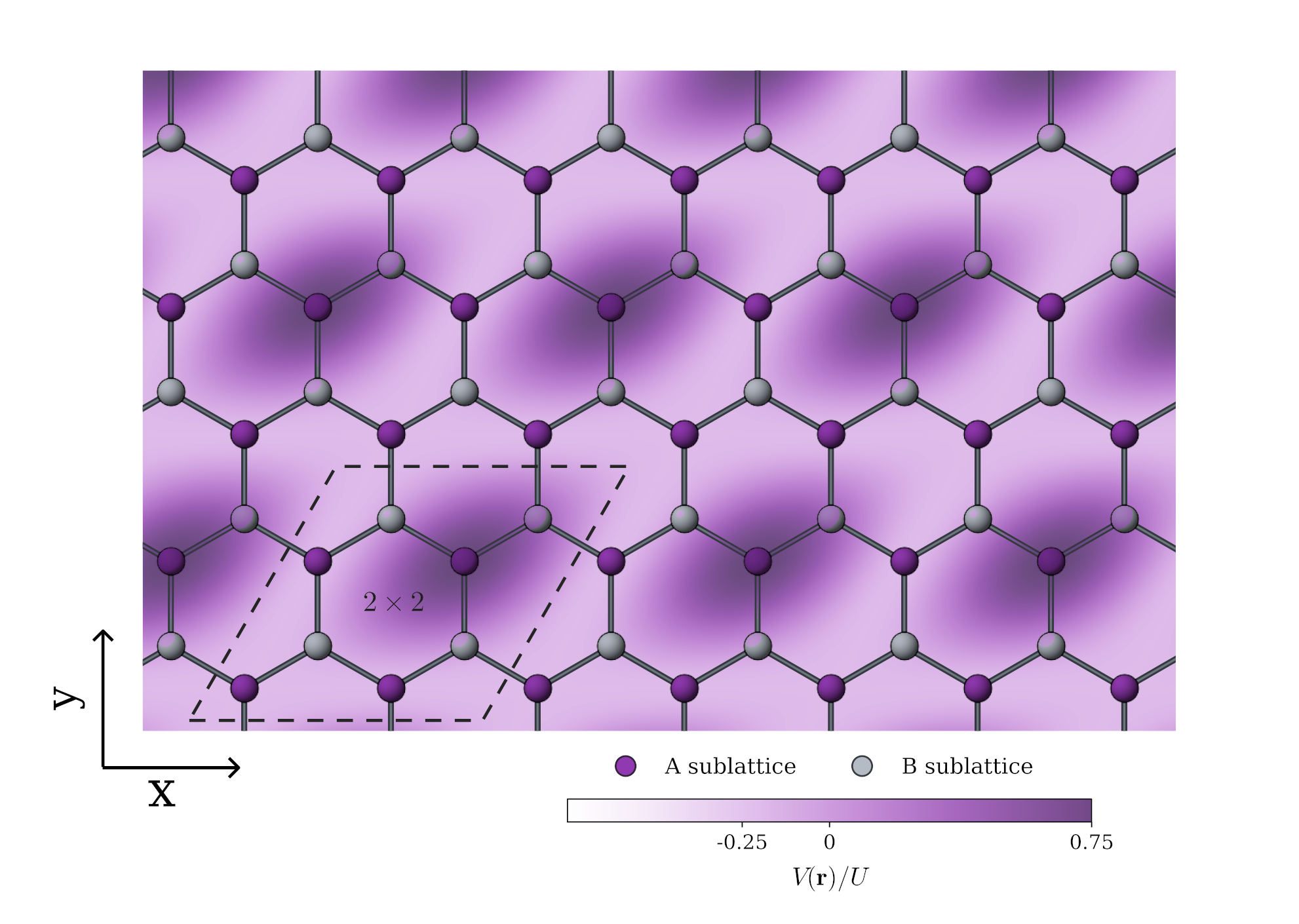}
\includegraphics[width=0.9\linewidth]{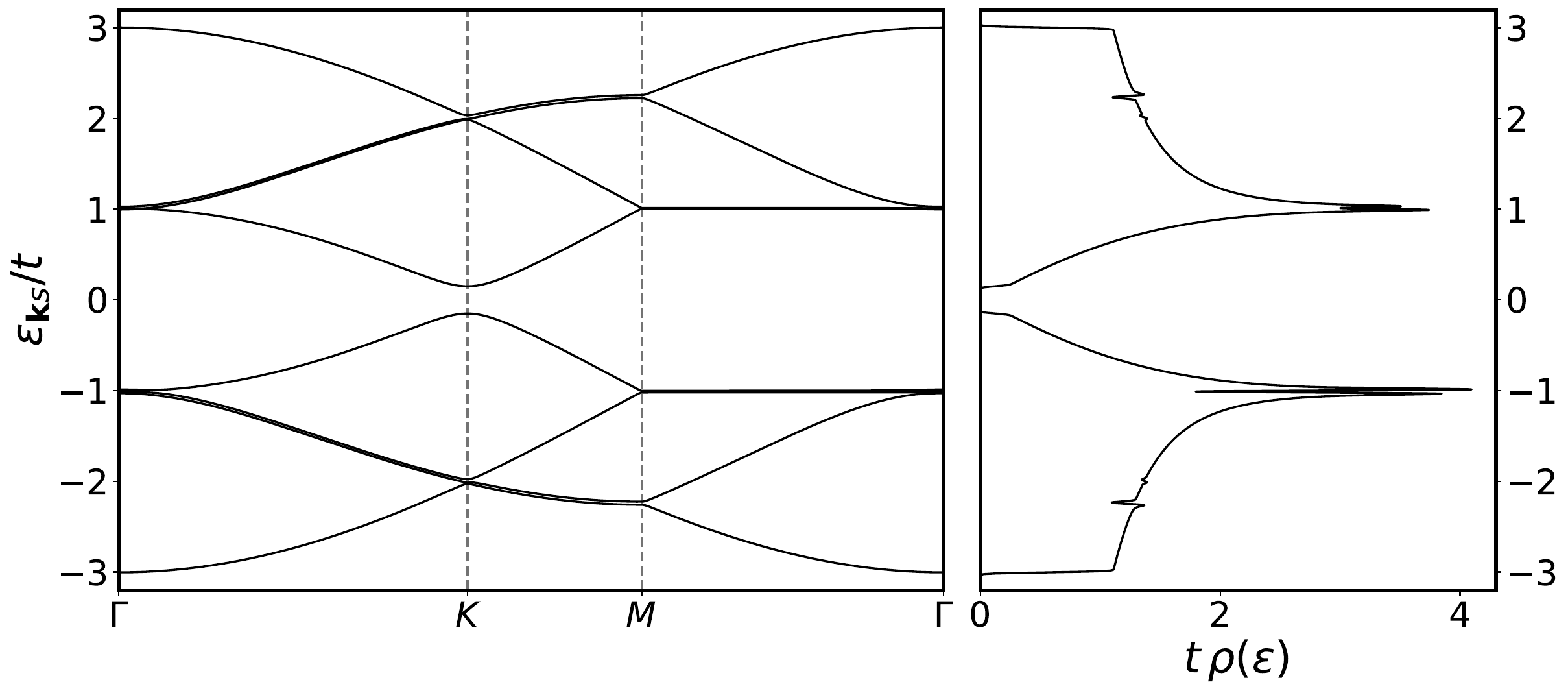}

\caption{Top: Graphene-like honeycomb lattice with a $2\times2$ supercell
and a periodic on-site potential. The general model also allows
a sublattice energy difference $\Delta$. The pristine-graphene
limit is obtained by setting $U=\Delta=0$.
Bottom left: Energy bands along the $\Gamma$--$K$--$M$--$\Gamma$ path. Bottom right: Density of states for the $2\times2$ supercell model. The density of states is normalized to $\int \rho(\epsilon)\,d\epsilon=8$ states per supercell. The values were obtained with $\Delta/t=0.3$ and $U/t=0.1$.}

\label{fig:graph}
\end{figure}
We evaluate the response for a nearest-neighbor tight-binding graphene-like model with a staggered sublattice potential and a periodic
on-site modulation,
\begin{align}
H_0=-t\!\sum_{\langle i,j\rangle}
\bigl(c_i^\dagger c_j+\mathrm{H.c.}\bigr)
+\sum_i\left[\frac{\Delta}{2}\eta_i+V(\vb r_i)\right]
 c_i^\dagger c_i.
\end{align}
Here $t$ is the nearest-neighbor hopping amplitude,
$\eta_i=+1$ ($-1$) on sublattice $A$ ($B$), and $\Delta$
sets the staggered sublattice energy difference. The periodic
on-site modulation is
\begin{equation}
V(\vb r)=U\sin^2\!\left(\frac{\vb r\cdot\vb b_1}{2}\right)
\sin^2\!\left(\frac{\vb r\cdot\vb b_2}{2}\right)-V_{\rm avg},
\label{eq:supercell_potential}
\end{equation}
where $U$ controls its strength, $\vb b_1$ and $\vb b_2$
are the reciprocal-lattice vectors of the $2\times2$ supercell, and
$V_{\rm avg}$ is chosen so that the average on-site potential
vanishes over the supercell. Periodic potentials are experimentally
realized in graphene--hexagonal-boron-nitride superlattices and can
reconstruct the electronic bands~\cite{yankowitz_emergence_2012}.
Such structures also exhibit photocurrent responses sensitive to
the symmetry and geometry of the minibands~\cite{delgado_notario_unveiling_2025}.
We use the $2\times2$ modulation as a minimal model of symmetry
breaking and band reconstruction, without attempting a quantitative
description of a moir\'e device.
Figure~\ref{fig:graph} illustrates the modulated lattice.

Unless stated otherwise, we use $\Delta=0.3t$, $U=0.1t$, and $\beta=1/(k_{\rm B}T)=1000/t$. The length $a_0$ used in the conductivity units denotes the carbon--carbon bond length. For $\Delta=U=0$, the model reduces to pristine graphene, for which the primitive two-atom unit cell suffices.

The electronic band structure and corresponding density of states for the default parameters are shown in the bottom part of Fig.~\ref{fig:graph}. The $2\times2$ supercell gives rise to eight bands, with an energy gap of approximately $\Delta$ separating the valence and conduction bands. This spectrum provides a reference for the optical transitions discussed below. In particular, $\mu=0$ lies within the gap, whereas $\mu=0.17t$, considered in the third-order response, lies near the conduction-band edge.

\subsection{Length- and velocity-gauge agreement with relaxation}
For the first benchmark, we set $\Delta=U=0$ and compare
our velocity-gauge calculation with the length-gauge relaxation
results of Passos \emph{et al.}~\cite{passos_nonlinear_2018}.
Both calculations include relaxation in the density-matrix equation
of motion,
Eq.~\eqref{eq:RTA}. We take $\mu=0.1t$ and
$f=f_{\rm FD}$ without equilibrium spectral broadening, while
retaining $\rho^{\rm eq}_{\vb k}(t)=f_{\rm FD}(H_{\vb k}(t))$.
\begin{figure}
  \centering
\includegraphics[width=\linewidth,height=0.4\textheight,keepaspectratio]{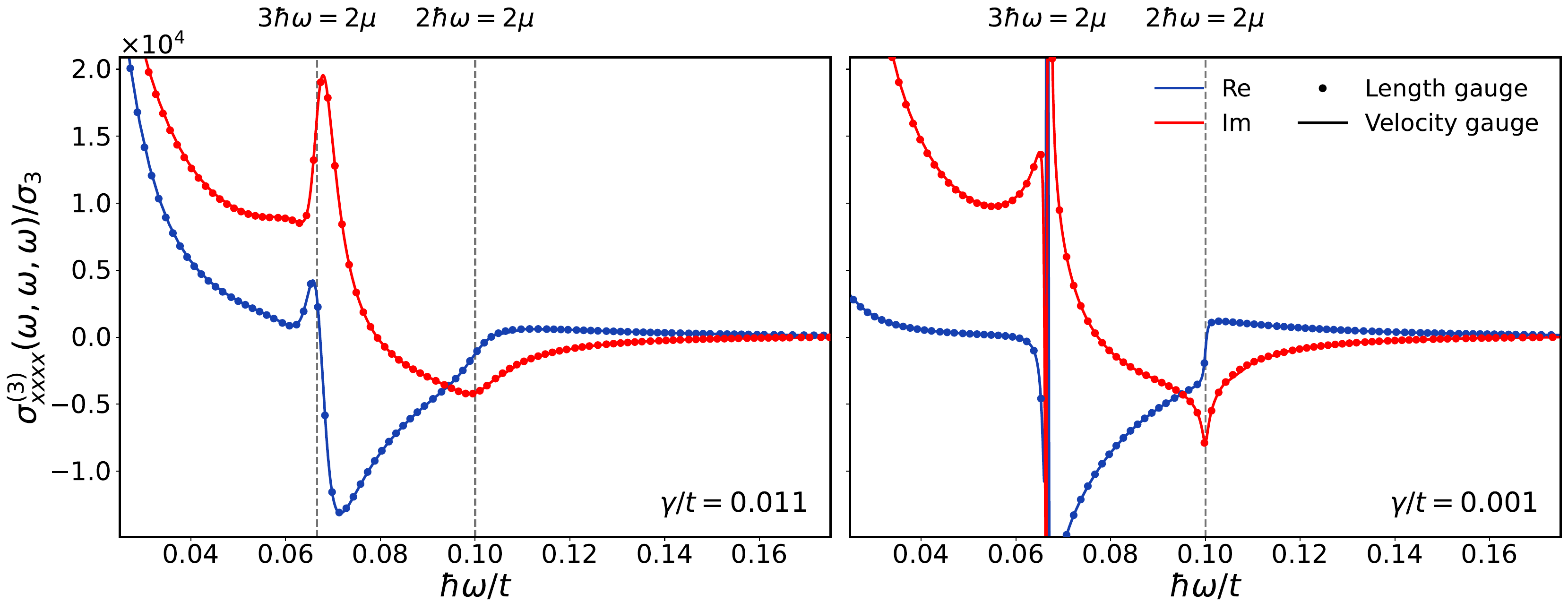}
  \caption{Third-harmonic conductivity
  $\sigma^{(3)}_{xxxx}(\omega,\omega,\omega)$ of pristine graphene.
  Circles: length-gauge relaxation data of
  Passos \emph{et al.}~\cite{passos_nonlinear_2018}; solid curves:
  present velocity-gauge result. Blue and red indicate the real and
  imaginary parts, respectively, for $\gamma/t=0.011$ (left) and
  $0.001$ (right), with $\mu/t=0.1$ and $\Delta=U=0$. Vertical dashed lines mark the three-photon ($3\hbar\omega=2\mu$) and two-photon ($2\hbar\omega=2\mu$) reference frequencies.
  The occupations are unbroadened, and
  $\sigma_3=e^4a_0^2/(8\hbar t^2)$. The length-gauge relaxation data was rescaled to our units and symmetrization convention.}
  \label{fig:thg_gauge_benchmark}
\end{figure}
Figure~\ref{fig:thg_gauge_benchmark} shows close agreement between the real
and imaginary parts of the two THG spectra,
including the three-photon resonance at $3\hbar\omega=2\mu$.
The benchmark extends the adiabatic-switching gauge comparison of
Ref.~\cite{passos_nonlinear_2018} to relaxation in the equation of
motion: the two gauges give the same response when the reference
equilibrium state is transformed consistently.

\subsection{Second-order optical response}
We now turn to the second-order response of gapped graphene with
periodic modulation. The same model introduced above describes
both second-harmonic generation, at output frequency $2\omega$,
and the dc photogalvanic response, at zero output frequency.
We compare the relaxation-time prescription of Eq.~\eqref{eq:RTA}
with adiabatic switching~\cite{ventura_gauge_2017,passos_nonlinear_2018}.
In the latter, the relaxation term is omitted from the equation of
motion, the equilibrium occupations are unbroadened, and each
input frequency is shifted as $\omega_i\to\omega_i+i\gamma/\hbar$.

\subsubsection{Second-harmonic generation}

We first consider the SHG conductivity $\sigma_{yyy}^{(2)}(\omega,\omega)$. For this calculation we take $\Delta=0.1t$, $U=0.1t$, and $\mu=\Delta/2$. The spectrum displays features near the two-photon ($2\hbar\omega\simeq\Delta$) and one-photon ($\hbar\omega\simeq\Delta$) interband resonances. Because the periodic modulation modifies the band structure, these frequencies serve as approximate reference scales.
\begin{figure}[h]
\centering
\includegraphics[width=\linewidth,keepaspectratio]{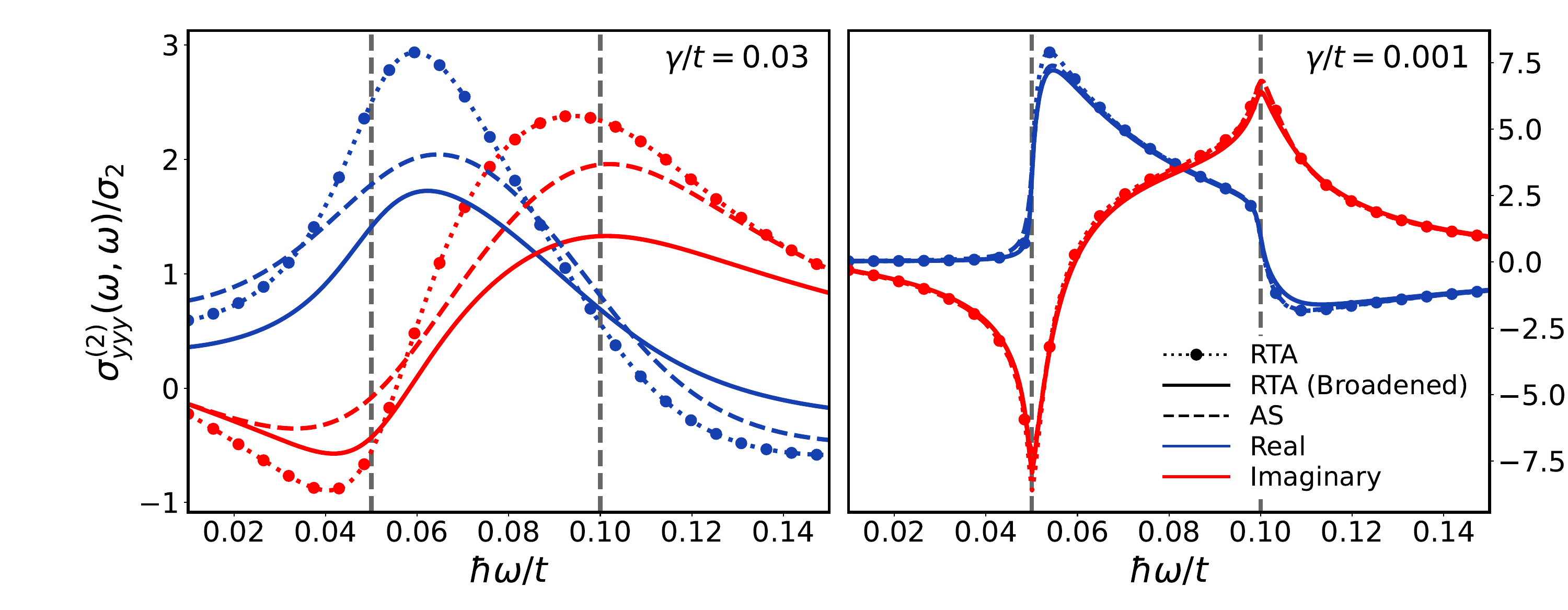}
  \caption{Real (blue) and imaginary (red) parts of the SHG
  conductivity $\sigma_{yyy}^{(2)}(\omega,\omega)$, calculated with
  relaxation (dotted lines with circular markers), relaxation with broadening (solid lines) and adiabatic switching (dashed lines).
  The left and right panels show $\gamma/t=0.03$ and $0.001$,
  respectively, for $\Delta=0.1t$, and $\mu=\Delta/2$.
  The spectral features occur near the two- and one-photon
  resonances, signaled by vertical lines at $\hbar\omega=\Delta/2$ and $\hbar\omega=\Delta$, respectively. Conductivities are given in units of
  $\sigma_2=e^3a_0/(4\hbar t)$~\cite{taghizadeh_linear_2017}.}
  \label{fig:SHG}
\end{figure}
Figure~\ref{fig:SHG} compares the real and imaginary parts of the SHG conductivity obtained using adiabatic switching and the relaxation-time approximation with either ordinary Fermi--Dirac or Lorentzian-broadened occupations.
At $\gamma/t=0.03$, the three calculations yield appreciably different peak amplitudes and line shapes, especially near the resonances.
Comparing the two relaxation-time results isolates the effect of equilibrium-occupation broadening while keeping the dynamical relaxation prescription fixed. In this case, the effect of broadening is the reduction of the magnitudes of the extrema in both the real and imaginary parts, while also modifying the shapes of the curves.
Relaxation rates of this order have been reported in both monolayer and bilayer graphene \cite{gnawali_high-order_2024,heide_electronic_2021,merboldt_observation_2025,mrudul_high-harmonic_2021}, indicating that relaxation-induced corrections to the response can be significant under experimentally relevant conditions. As $\gamma\rightarrow 0$, agreement between the three calculations improves. At $\gamma/t=0.001$, the spectra nearly coincide over most of the frequency range, with residual differences concentrated near the one- and two-photon resonances.
This comparison therefore highlights the sensitivity of the finite-$\gamma$ SHG response to both the dynamical relaxation prescription and the broadening of equilibrium occupations.

\subsubsection{DC photogalvanic response}

We next consider the dc conductivity
$\sigma_{yyx}^{(2)}(\omega,-\omega)$, returning to the default
parameters $\Delta=0.3t$ and $U=0.1t$, with $\mu=0$. In noncentrosymmetric
systems, the dc photogalvanic response can contain a shift-current
contribution, associated with charge displacement during optical
transitions, and an injection-current contribution, associated
with asymmetric populations of photoexcited carriers
\cite{sipe_second-order_2000,morimoto_topological_2016,de_juan_quantized_2017}.
The supercell modulation lowers the symmetry of uniformly gapped
graphene, allowing additional second-order tensor components.
For this tensor component, the real part describes the shift-type
response, while the imaginary part describes the helicity-dependent
channel containing the injection contribution.

\begin{figure}[h]
  \centering
  \includegraphics[width=\linewidth,height=0.72\textheight,keepaspectratio]{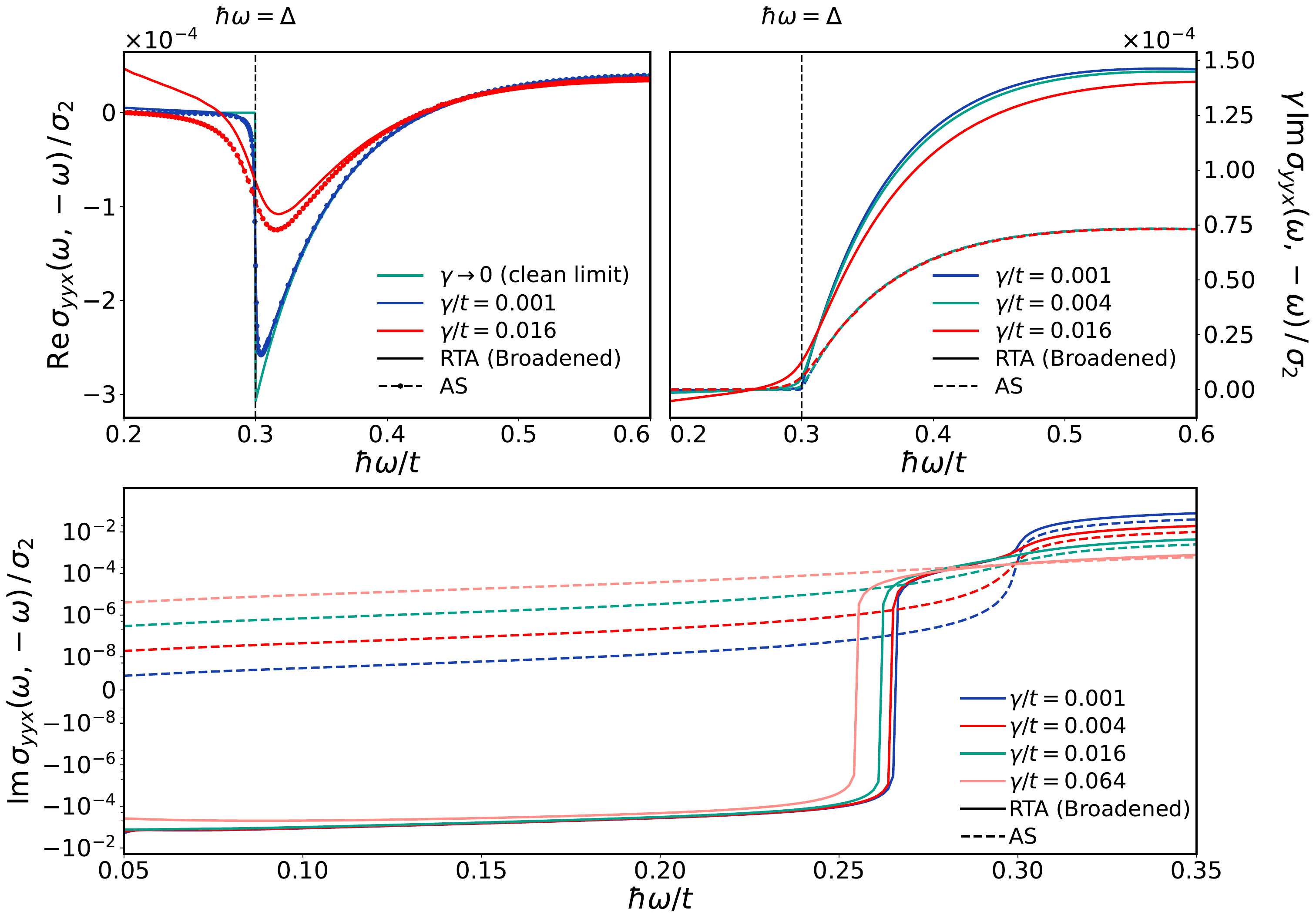}
  \caption{Second-order dc conductivity
  $\sigma_{yyx}^{(2)}(\omega,-\omega)$ for relaxation through the
  density-matrix equation of motion (solid lines) and adiabatic
  switching (dashed lines). Upper left panel: real part for
  $\gamma/t=0.001$ and $0.016$, including a clean-limit reference.
  Upper right panel: $\gamma\,\Im\sigma_{yyx}^{(2)}$ for
  $\gamma/t=0.001$, $0.004$, and $0.016$; the rescaled
  curves above the gap are consistent with leading $1/\gamma$ and
  $1/(2\gamma)$ scalings for relaxation and adiabatic switching,
  respectively. Bottom panel: imaginary part below the gap for
  $\gamma/t=0.001$, $0.004$, $0.016$, and $0.064$, shown on a
  symmetric logarithmic scale. The subgap curves indicate a finite, nonzero small-$\gamma$ limit
  for relaxation and a vanishing response for adiabatic switching. The vertical
  line marks the approximate optical threshold $\hbar\omega=\Delta$;
  here $\Delta=0.3t$, $U=0.1t$, and $\mu=0$. Conductivities are given in
  units of $\sigma_2=e^3a_0/(4\hbar t)$.}
  \label{fig:shift_injection_3panel}
\end{figure}

Figure~\ref{fig:shift_injection_3panel} summarizes the dependence
of the dc response on frequency and relaxation strength.
The upper left panel shows the real part of the conductivity, associated
with the shift-current contribution. A pronounced negative feature
develops near the approximate optical threshold $\hbar\omega=\Delta$.
At finite $\gamma$, the relaxation-time and adiabatic-switching
prescriptions broaden this structure differently. As $\gamma\to0^+$,
both approach the same finite response away from the threshold,
as shown by comparison with the clean-limit reference curve.

The upper right panel shows $\gamma\,\Im\sigma_{yyx}^{(2)}$, which isolates the leading inverse-relaxation contribution. Above the optical threshold, the rescaled quantity $\gamma\Im\sigma$ approaches a finite limit as $\gamma$ is decreased, consistent with an injection-current contribution. In this frequency region, away from the immediate vicinity of the threshold, the leading terms obey
\begin{align}
\Im\sigma_{yyx,\mathrm{RTA}}^{(2)}(\omega,-\omega;\gamma)
&\sim\frac{I_{yyx}(\omega)}{\gamma},\\
\Im\sigma_{yyx,\mathrm{AS}}^{(2)}(\omega,-\omega;\gamma)
&\sim\frac{I_{yyx}(\omega)}{2\gamma},
\qquad \hbar\omega>\Delta,
\label{eq:dc_injection_scaling}
\end{align}
where $I_{yyx}(\omega)$ is the limiting injection coefficient
in this comparison. Thus, at the same numerical value of $\gamma$,
the leading relaxation-time contribution is approximately twice
the adiabatic-switching one. The factor of two follows from the
different dc denominators: the second-order density-matrix
response contains $i\gamma$ in the relaxation-time prescription,
whereas adiabatic switching gives $2i\gamma$ after shifting both
input frequencies~\cite{passos_nonlinear_2018}. The factor of two therefore refers to the leading above-gap term
at equal $\gamma$.

The lower panel resolves the imaginary part below the optical gap,
where the leading $1/\gamma$ injection divergence is absent.
The numerical results indicate a finite, nonzero small-$\gamma$
limit for the relaxation-time response and a vanishing limit for
adiabatic switching. For the subgap frequencies shown, this
behavior is summarized by
\begin{align}
\lim_{\gamma\to0^+}
\Im\sigma_{yyx,\mathrm{RTA}}^{(2)}(\omega,-\omega;\gamma)
&=B_{yyx}(\omega),\\
\lim_{\gamma\to0^+}
\Im\sigma_{yyx,\mathrm{AS}}^{(2)}(\omega,-\omega;\gamma)
&=0,
\label{eq:subgap_limit}
\end{align}
with $B_{yyx}(\omega)$ generally different from zero. This finite subgap term is distinct from the divergent above-gap
injection contribution. The imaginary parts thus indicate different pointwise small-$\gamma$ limits, even where the real parts converge to the same clean-limit response.

\FloatBarrier
\subsection{Third-order optical response}

We finally turn to third-harmonic generation (THG), described
by $\sigma_{xxxx}^{(3)}(\omega,\omega,\omega)$. Unlike the preceding
second-order calculations, we place the chemical potential in
the conduction band at $\mu=0.17t$ to investigate the effects
of relaxation and equilibrium broadening for a partially
occupied band. With $\Delta=0.3t$, this is slightly above the
unmodulated conduction-band minimum $\Delta/2=0.15t$.

\begin{figure}[!htbp]
	\centering
	\includegraphics[width=\linewidth,height=0.4\textheight,keepaspectratio]{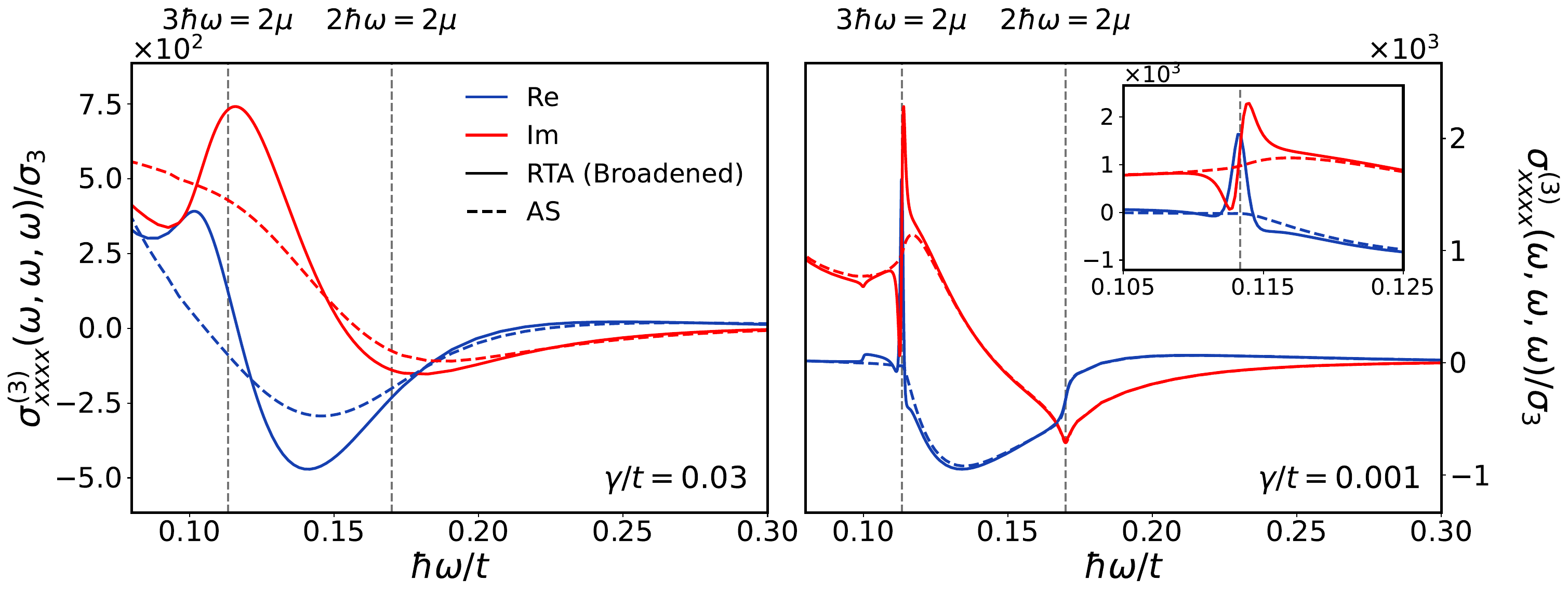}
	\caption{Real (blue) and imaginary (red) parts of the third-harmonic generation conductivity, $\sigma_{xxxx}^{(3)}(\omega,\omega,\omega)$, calculated with relaxation and broadened occupations (solid lines) and adiabatic switching (dashed lines), for $\gamma=0.03t$ (left panel) and $\gamma=0.001t$ (right panel). The staggered potential and the chemical potential are fixed at $\Delta=0.3t$ and $\mu=0.17t$, respectively. Vertical dashed lines mark the three-photon ($3\hbar\omega=2\mu$) and two-photon ($2\hbar\omega=2\mu$) reference frequencies. The inset resolves the three-photon feature, where a marked difference remains at $\gamma/t=0.001$. Conductivities are expressed in units of $\sigma_3=\frac{e^4a_0^2}{8\hbar t^2}$ \cite{ventura_study_2020}.}
	\label{fig:thg}
\end{figure}

Figure~\ref{fig:thg} compares the THG spectra for
$\gamma/t=0.03$ and $0.001$. At the larger value of $\gamma$, the two
prescriptions give appreciably different amplitudes and line shapes.
Reducing $\gamma$ improves agreement over most of the spectrum,
but a substantial difference remains near the three-photon feature
at $3\hbar\omega\simeq2\mu$. The inset resolves this narrow region,
where the relaxation-time result has a sharp structure in both
parts of the conductivity. Agreement is closer near the two-photon
feature at $2\hbar\omega\simeq2\mu$.
These frequencies provide reference scales for the modulated system.
The comparison shows that small damping alone does not ensure
agreement near multiphoton resonances.

\input{conclusions}

\funding


The authors acknowledge support from the Portuguese Foundation for Science and Technology (FCT-Portugal) in the framework of the Strategic Funding UID/04650/2025, Grant No. UID/PRR/04650/2025 (DOI: https://sciproj.ptcris.pt/177318UID/), and Grant No. UID/PRR2/04650/2025 (DOI: https://sciproj.ptcris.pt/178170UID/).   J.P., B.A. and J.M.V.P.L also acknowledge support from FCT-Portugal through the project QPI-MoireS (DOI: https://sciproj.ptcris.pt/177781PRJ/). Finally, there has been support from the Erasmus+: Erasmus
Mundus programme of the European Union.

\appendix
\section{Derivation of the Equilibrium Density Matrix Expansion}\label{app:deriv_expansion}
We first assume a power series $f(x)=\sum_{n\ge 0}a_nx^n$
convergent on the spectrum. Thus the equilibrium density matrix is
\begin{equation}
\rho^{\rm eq}_{\vb k}(t)
=\sum_{n=0}^{\infty}a_n
\bigl(H_{0,\vb k}+V_{\vb k}(t)\bigr)^n.
\end{equation}
As $H_{0,\vb k}$ and $V_{\vb k}$ need not commute, each power of $H_{0,\vb k}+V_{\vb k}(t)$ can be organized according
to the number of insertions of the perturbation $V_{\vb k}(t)$. Suppressing the time and $\vb k$ dependence in the Hamiltonian for compactness:
\begin{equation}
(H_0+V)^n=H_0^n+\sum_{l=1}^{n}S_l^{\,n-l}.
\end{equation}
Here, $S_l^{N}$ denotes the sum of all operator products containing
exactly $l$ factors of $V$ and $N$ factors of $H_0$.  Accordingly, the
equilibrium density matrix admits the perturbative expansion
\begin{equation}
\rho^{\rm eq}_{\vb k}(t)
=\sum_{l=0}^{\infty}\rho^{[l]}_{\vb k}(t),
\end{equation}
where $\rho^{[l]}_{\vb k}$ likewise collects all terms containing $l$ insertions of $V$,
not field order, since $V$ contains all powers of $\vb A(t)$.  In the eigenbasis $H_{0,\vb k}\ket{s}=\epsilon_{\vb ks}\ket{s}$, the $l\ge1$ insertion terms obey
\begin{equation}
\mel{s_0}{S_l^N}{s_l}=\sum_{s_1,\ldots,s_{l-1}}
 V_{s_0s_1}\cdots V_{s_{l-1}s_l}
c_l^N(\epsilon_{s_0},\ldots,\epsilon_{s_l}),
\end{equation}
with the symmetric polynomial $c_l^N(\epsilon_{s_0},\ldots,\epsilon_{s_l})$ given by
\begin{align}
c_l^N(\epsilon_{s_0},\ldots,\epsilon_{s_l})=
\sum_{n_0+\cdots+n_l=N}
\epsilon_{s_0}^{n_0}\cdots
\epsilon_{s_l}^{n_l}.
\label{eq:little_s}
\end{align}
This leads us to the form used in Equation \eqref{eq:expansion_potentialV}, repeated below:

\begin{equation*}
\rho^{[l]}_{\vb k,s_0s_l}(t)=\sum_{s_1,\ldots,s_{l-1}}
 V_{\vb k,s_0s_1}(t)\cdots
 V_{\vb k,s_{l-1}s_l}(t)\,
\mathcal C_{s_0\cdots s_l},
\end{equation*}
with the coefficient $\mathcal C_{s_0\cdots s_l}$ defined as:
\begin{equation}
\mathcal C_{s_0\cdots s_l}
=\sum_{n=l}^{\infty}a_n\,
 c_l^{\,n-l}(\epsilon_{s_0},\ldots,\epsilon_{s_l}).
\label{c_coefs}
\end{equation}

As it stands, the constrained nature of the summation in Equation
(\ref{eq:little_s}) is not very practical for our purposes. Instead,
one can prove that
(see Appendix~\ref{app:coefficients}) 
\begin{equation}
c_l^N(\epsilon_{s_0},\ldots,\epsilon_{s_l})
=\sum_{j=0}^{l}
\frac{\epsilon_{s_j}^{\,N+l}}
 {\displaystyle\prod_{\ell\ne j}
 (\epsilon_{s_j}-\epsilon_{s_\ell})}.
\label{eq:little_s_melhor}
\end{equation}
Substituting Eq.~\eqref{eq:little_s_melhor} into
Eq.~\eqref{c_coefs} and using the definition $\sum_{n=0} a_n \varepsilon ^n = f(\varepsilon)$ allows us to write
\begin{align}
\mathcal C_{s_0\cdots s_l}
=\sum_{j=0}^{l}
\frac{f(\epsilon_{s_j})
-\displaystyle\sum_{r=0}^{l-1}a_r\epsilon_{s_j}^{\,r}}
 {\displaystyle\prod_{\ell\ne j}
 (\epsilon_{s_j}-\epsilon_{s_\ell})}.
\end{align}
The second contribution vanishes identically due to the Lagrange interpolation
identity (proved in Appendix~\ref{app:lagrange}),
\begin{equation}
\sum_{j=0}^{l}
\frac{\epsilon_{s_j}^{\,r}}
 {\displaystyle\prod_{\ell\ne j}
 (\epsilon_{s_j}-\epsilon_{s_\ell})}
=\delta_{rl},\qquad 0\le r\le l.
\label{eq:lagrange_main}
\end{equation}
leaving the remarkably compact result shown in Equation \eqref{eq:large_S}:
\begin{equation*}
\mathcal C_{s_0\cdots s_l}
=\sum_{j=0}^{l}
\frac{f(\epsilon_{s_j})}
 {\displaystyle\prod_{\ell\ne j}
 (\epsilon_{s_j}-\epsilon_{s_\ell})}.
\end{equation*}

\section{Algebraic identities}

\subsection{Derivation of the coefficients}
\label{app:coefficients}

The polynomial $c_l^N$ in Eq.~\eqref{eq:little_s} is the coefficient
of $x^N$ in the generating function
\begin{equation}
Z(x)=\prod_{j=0}^{l}\frac{1}{1-\epsilon_{s_j}x}.
\label{eq:partition_function}
\end{equation}
Expanding each factor around $x=0$ gives
\begin{align}
Z(x)&=\prod_{j=0}^{l}\sum_{n_j=0}^{\infty}(\epsilon_{s_j}x)^{n_j}
\nonumber\\
&=\sum_{N=0}^{\infty}x^N
\sum_{n_0+\cdots+n_l=N}\prod_{j=0}^{l}\epsilon_{s_j}^{n_j}.
\end{align}
For distinct energies, the partial-fraction decomposition is
\begin{equation}
Z(x)=\sum_{j=0}^{l}\frac{C_j}{1-\epsilon_{s_j}x},
\label{eq:prod_to_sum}
\end{equation}
with
\begin{equation}
C_j=\frac{\epsilon_{s_j}^{l}}
{\displaystyle\prod_{\ell\ne j}(\epsilon_{s_j}-\epsilon_{s_\ell})}.
\label{eq:Coeffs_partial_fraction}
\end{equation}
These coefficients follow by multiplying Eq.~\eqref{eq:prod_to_sum}
by $1-\epsilon_{s_j}x$ and evaluating at $x=1/\epsilon_{s_j}$;
zero energies are included by continuity. Hence
\begin{equation}
Z(x)=\sum_{j=0}^{l}\frac{1}{1-\epsilon_{s_j}x}
\frac{\epsilon_{s_j}^{l}}
{\displaystyle\prod_{\ell\ne j}(\epsilon_{s_j}-\epsilon_{s_\ell})}.
\label{eq:coiso}
\end{equation}
Expanding once more,
\begin{equation}
Z(x)=\sum_{N=0}^{\infty}x^N\sum_{j=0}^{l}
\frac{\epsilon_{s_j}^{N+l}}
{\displaystyle\prod_{\ell\ne j}(\epsilon_{s_j}-\epsilon_{s_\ell})},
\end{equation}
which establishes Eq.~\eqref{eq:little_s_melhor}.

\subsection{Lagrange interpolation identity}
\label{app:lagrange}

We now prove the identity used in Appendix \ref{app:deriv_expansion},
\begin{align}
\sum_{j=0}^{l}
\frac{\epsilon_{s_j}^{\,r}}
{\displaystyle\prod_{\ell\neq j}
(\epsilon_{s_j}-\epsilon_{s_\ell})}
=
\delta_{rl},
\qquad 0\leq r\leq l,
\label{eq:lagrange_identity}
\end{align}
for distinct energies $\epsilon_{s_j}$.

Consider a polynomial $p(x)$ of degree at most $l$. The Lagrange
interpolation formula allows it to be written as
\begin{align}
p(x)
=
\sum_{j=0}^{l}
p(\epsilon_{s_j})\,L_j(x),
\end{align}
where
\begin{align}
L_j(x)
=
\prod_{\ell\neq j}
\frac{x-\epsilon_{s_\ell}}
{\epsilon_{s_j}-\epsilon_{s_\ell}}
\end{align}
are the Lagrange basis polynomials. The coefficient of $x^l$ in each basis polynomial is
\begin{align}
[x^l]\,L_j(x)
=
\frac{1}
{\displaystyle\prod_{\ell\neq j}
(\epsilon_{s_j}-\epsilon_{s_\ell})},
\end{align}
where $[x^l]$ denotes the coefficient of $x^l$.

Choosing
\begin{align}
p(x)=x^r,
\qquad 0\leq r\leq l,
\end{align}
and comparing the coefficients of $x^l$ on both sides of the
interpolation formula gives
\begin{align}
[x^l]\,x^r
=
\sum_{j=0}^{l}
\frac{\epsilon_{s_j}^{\,r}}
{\displaystyle\prod_{\ell\neq j}
(\epsilon_{s_j}-\epsilon_{s_\ell})}.
\end{align}
The coefficient on the left-hand side vanishes for $r<l$ and is equal
to unity for $r=l$. Therefore,
\begin{align}
\sum_{j=0}^{l}
\frac{\epsilon_{s_j}^{\,r}}
{\displaystyle\prod_{\ell\neq j}
(\epsilon_{s_j}-\epsilon_{s_\ell})}
=
\delta_{rl},
\qquad 0\leq r\leq l,
\end{align}
which establishes the identity used in the derivation of the density matrix expansion coefficients.

\bibliographystyle{iopart-num}
\bibliography{NonLinearOptics}

\end{document}

%% file: introduction.tex
\section{Introduction}

The study of nonlinear optical responses has become fundamental in understanding and characterizing the electronic properties of crystalline materials. In fact, ever since second harmonic generation was first observed~\cite{franken_generation_1961}, frequency conversion and wave mixing have become important methods to probe solids and interfaces~\cite{bloembergen_nonlinear_1982,shen_surface_1989}. In graphene, coherent four-wave mixing~\cite{hendry_coherent_2010}, third-harmonic generation~\cite{kumar_third_2013}, and electrical tuning of nonlinear resonances~\cite{soavi_broadband_2018,jiang_gate-tunable_2018,jiang_author_2018} have been observed. Additionally, there have been developments in high-harmonic generation in solids~\cite{ghimire_high-harmonic_2019}, anisotropic second-harmonic generation~\cite{wu_giant_2017}, and spatially dispersive photogalvanic currents in Weyl semimetals~\cite{ji_spatially_2019}.

Semiclassical band dynamics provides useful intuition for intraband motion, while interband transitions and coherences require a quantum description, naturally formulated within a single-particle reduced density-matrix, $\rho(t)$~\cite{ghimire_high-harmonic_2019} . Light-matter coupling can be described with different gauge choices, and for a homogeneous electric field it is common to employ either the  length gauge (coupling via a  $\mathbf{r}\cdot\mathbf{E}$ scalar potential, with $\mathbf{r}$ the position operator and $\mathbf{E}$ the electric field) or the velocity gauge  (coupling via the $\mathbf{p} \rightarrow \mathbf{p} -q\mathbf{A}$ minimal substitution, with $\mathbf{p}$ the momentum operator, $\mathbf{A}$ the vector potential and $q$ the particle charge) ~\cite{sipe_nonlinear_1993,aversa_third-order_1994,aversa_nonlinear_1995,passos_nonlinear_2018}. Formal approaches to obtain second-order responses have been developed~\cite{hughes_calculation_1996,sipe_second-order_2000}, as well as numerical length-gauge density-matrix methods~\cite{virk_semiconductor_2007}. For graphene, there have been studies addressing intraband and interband processes and how they lead to non-linear behavior~\cite{mikhailov_non-linear_2007}. Moreover, graphene has been extensively studied in the context of doping and relaxation~\cite{glazov_high_2014,cheng_third_2014,cheng_third-order_2015,mikhailov_quantum_2016}, photocurrents~\cite{hipolito_nonlinear_2016}, high harmonics in extended and confined geometries~\cite{al-naib_high_2014,mrudul_high-harmonic_2021,gnawali_high-order_2024}, and plasmon-assisted second-harmonic generation~\cite{pinho_plasmon_2025}.

The length and velocity gauges provide equivalent descriptions of optical response when the light--matter coupling is treated consistently, and their respective computational properties are well established~\cite{aversa_nonlinear_1995,ventura_gauge_2017,passos_nonlinear_2018}. Here we adopt the velocity gauge, in which a spatially uniform electromagnetic field is described by a time-dependent vector potential, $\mathbf{A}(t)$, with $\mathbf{E}(t)=-\partial_t\mathbf{A}(t)$. Unlike the length gauge, where the position operator introduces momentum derivatives acting on the density matrix, the velocity gauge preserves translational symmetry and allows the coherent dynamics to be expressed in terms of matrices that are local in momentum space. This considerably simplifies the perturbative expansion of the density matrix. In the presence of relaxation, however, the field dependence of the reference equilibrium state can reintroduce momentum derivatives of the density matrix into the calculation \cite{passos_nonlinear_2018}. As we show in this work, these derivatives can be avoided within the relaxation-time approximation, independently of the microscopic details of the single-particle Hamiltonian.

When simulating optical response of solids, it is important to include dissipative effects, which relax the system towards thermodynamic equilibrium. The simplest approach in which this is taken into account, within perturbation theory, is via adiabatic switching (AS), where the time dependent perturbation is assumed  to have the form $\mathbf{A}(t) \propto e^{-i(\omega + i \gamma )t}$ , where $\gamma$ describes the slow switching-on of the external field and effectively acts as a broadening term, mocking dissipation. Another, more justifiable and commonly used, approach is the relaxation-time approximation (RTA), in which a term of the form $ - \gamma \left( \rho(t) - \rho^{\text{eq}}(t) \right) \subset \partial_t \rho(t)$, is included in the equation of motion for the density matrix $\rho(t)$, with $\rho^\text{eq}(t)$ being an equilibrium density matrix, towards which the system tends to relax. It has been pointed out that the AS and RTA do not, in general, coincide~\cite{cheng_third-order_2015,passos_nonlinear_2018,pinho_exact_2026}. In the latter approach, and in the velocity gauge, the equilibrium state towards which the system relaxes must follow the applied field and therefore acquires an explicit dependence on the vector potential in the velocity gauge. It is worth pointing out that the RTA has been justified for systems where dissipation is mocked via hybridization with fermionic baths, Bütikker probes~\cite{pinho_exact_2026}. Coupling to these baths not only modifies the dynamics of the density matrix, but also changes the equilibrium state. Hence, the choice of relaxation mechanism employed in the dynamics of the system can play a significant role for stationary injection currents and resonant nonlinear responses.

The treatment of relaxation matters when nonlinear optics probes quantum geometry and topology~\cite{morimoto_topological_2016,orenstein_topology_2021,jiang_revealing_2025}. The predicted quantization of the circular photogalvanic injection-rate trace under suitable conditions~\cite{de_juan_quantized_2017}, and proposals to detect topology through high-harmonic generation~\cite{silva_topological_2019}, connect nonlinear signals to Bloch-state structure. A consistent treatment of relaxation is therefore essential for disentangling intrinsic contributions rooted in the quantum geometry of Bloch states from extrinsic, lifetime-dependent contributions arising from the nonequilibrium carrier distribution.

Here we incorporate relaxation in the equation of motion, for the density matrix, via the relaxation-time approximation (RTA), within the velocity gauge , using a reference equilibrium state that adiabatically follows the applied field. Compared with previous formulations based on the length gauge, the present formulation avoids cumbersome derivatives of the density matrix with respect to momentum. To illustrate the formalism, we consider models for which minimal coupling is implemented through a shift of crystal momentum, $\mathbf{k}\rightarrow\mathbf{k}-q\mathbf{A}/\hbar$, allowing the perturbative coefficients of the Hamiltonian to be obtained directly from momentum derivatives of the unperturbed Bloch Hamiltonian. This choice simplifies the applications presented here, but is not a requirement of the general framework. We obtain recursive nonlinear-conductivity expressions to arbitrary order and a closed formula for the field expansion of the equilibrium density matrix in terms of divided differences of the occupation function. The latter accommodates broadened occupations, induced by hybridization with fermionic baths. 

We apply the framework to a gapped honeycomb model, subject to a periodic potential, computing: second-harmonic generation, dc shift and injection currents, and third-harmonic generation. At finite damping, the results can differ markedly from adiabatic switching: at equal $\gamma$, the injection-current response has a $1/\gamma$ contribution with an approximately twofold larger prefactor in the relaxation-time scheme, while discrepancies persist near multiphoton resonances in the third-order response even for small damping.

Section~2 introduces the formalism, Sec.~3 derives the equilibrium density matrix expansion, Sec.~4 presents the numerical results, and Sec.~5 summarizes our conclusions.

%% file: conclusions.tex
\section{Conclusion}
\label{sec:conclusion}

We have derived a velocity-gauge formulation of nonlinear optical
response with relaxation towards a field-dependent equilibrium
state. The conductivity follows recursively from the density-matrix
equation of motion, while the equilibrium-density $V_{\vb k}$-expansion coefficients are
given in terms of the unperturbed spectra and the occupation function. This
provides a calculation at arbitrary order, includes bath-broadened occupations, and requires no
derivatives of the density matrix. Additionally, for tight-binding or continuum Hamiltonians, derivatives of Bloch eigenvectors are likewise avoided.

The graphene calculations establish two distinct results. First,
length- and velocity-gauge conductivities agree when relaxation is
treated consistently. Second, this gauge equivalence does not imply
equivalence between relaxation and adiabatic switching. The latter
two prescriptions give different harmonic-generation spectra at
finite $\gamma$, with discrepancies still resolved near the
three-photon resonance at the smallest $\gamma$ considered.

The dc response makes the distinction particularly clear. Away from
the optical threshold, the shift-current component approaches the
same finite limit in both prescriptions. The leading above-gap
injection contribution instead scales as $1/\gamma$ for relaxation
and $1/(2\gamma)$ for adiabatic switching. This factor of two refers
to the same numerical value of $\gamma$ and follows from the
respective dc denominators. Below the gap, the numerical results indicate a finite, nonzero
imaginary response in the small-$\gamma$ limit of the relaxation-time
calculation, whereas the adiabatic-switching result vanishes. Thus the difference extends
beyond resonance broadening to the limiting dc response itself.

A relaxation prescription therefore forms part of the physical
model of nonlinear response. Adiabatic switching cannot in general
replace a relaxation term, and a small broadening parameter alone
does not ensure that the two calculations agree. The present
formulation makes the field dependence of the relaxation state
explicit and provides a direct way to retain it in multiband
velocity-gauge calculations.